\documentstyle[sprocl,epsfig,axodraw]{article}

\def\lsim{\mathrel{\raise.3ex\hbox{$<$\kern-.75em\lower1ex\hbox{$\sim$}}}}
\def\gsim{\mathrel{\raise.3ex\hbox{$>$\kern-.75em\lower1ex\hbox{$\sim$}}}}

\def\PRD{{\em Phys. Rev.} D }
\def\PLB{{\em Phys. Lett.}  B }
\def\PLBB{{\em Phys. Lett.}  B}
\def\NIMA{{\em Nucl. Instrum. Methods} A }

\newcommand{\beq}{\begin{eqnarray}}
\newcommand{\eeq}{\end{eqnarray}}

\begin{document}


\title{$\tau\tau$ Fusion to SUSY Higgs Bosons at a Photon Collider:\\[2mm]
       Measuring \boldmath{$\tan\beta$}}

\author{S.~Y.~Choi}

\address{Dept.\ Physics, Chonbuk National University, 
               Chonju 561--756, Korea}

\author{J.~Kalinowski}

\address{Inst.\ Theor.\ Physics, Warsaw University, PL--00681 Warsaw, 
               Poland}

\author{J.~S.~Lee}

\address{Dept.\ Physics and Astronomy, Univ.\  Manchester, 
               Manchester M13 9PL, UK}

\author{M.~M.~M\"uhlleitner, M.~Spira}

\address{Paul Scherrer Institut, CH-5232 Villigen PSI, Switzerland}

\author{P.~M.~Zerwas}

\address{Deutsches Elektronen--Synchrotron DESY, D--22603 Hamburg, 
Germany}

\maketitle\abstracts{
$\tau\tau$ fusion to light $h$ and heavy $H,A$ Higgs bosons is 
investigated in the Minimal Supersymmetric Standard Model (MSSM) at a photon 
collider as a promising channel for measuring large values of $\tan\beta$. 
For standard design parameters of a photon collider an error 
$\Delta\tan\beta\sim 1$, uniform for $\tan\beta \gsim 10$, may be expected, 
improving on  complementary measurements at LHC and $e^+e^-$ linear 
colliders.
}%

{\bf 1.)} The measurement of the mixing parameter $\tan\beta$, 
one of the fundamental parameters in the Higgs sector of the
Minimal Supersymmetric Standard Model [MSSM] and other supersymmetry
scenarios, is a difficult task. Many of the observables, in the
chargino/neutralino sector for instance, involve only $\cos2 \beta$
and thus are quite insensitive to the parameter $\tan\beta$ 
for large values.
Remarkably different however are the heavy $H/A$ Higgs couplings to down-type
fermions which, for values of the pseudoscalar Higgs boson mass at the
electroweak scale and beyond, both are directly proportional to $\tan\beta$ if
this parameter becomes large, see {\it e.g.} Ref.~\cite{R2}, so that
they are highly sensitive to its value. Also the down-type couplings 
of the light
$h$ Higgs boson in the MSSM are close to $\tan\beta$ if the pseudoscalar mass
is moderately small.

In this note we point out that $\tau\tau$ fusion to Higgs bosons at a photon
collider~\cite{2a} can provide a valuable method for measuring 
$\tan\beta$,
after searching for and exploring Higgs bosons in $\gamma\gamma$ 
fusion~\cite{2b,2c}. 
The entire Higgs mass range up to the kinematical limit  
can be covered for large $\tan\beta$ by this method.    

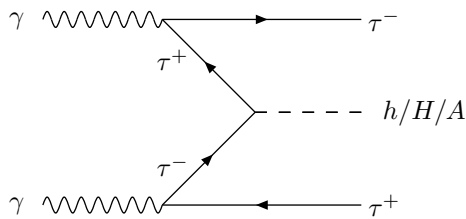
\begin{figure}[htb]
\begin{center}
\begin{picture}(100,100)(240,0)
\Text(221,85)[]{$\gamma$}
\Photon(230,85)(275,85){3}{7}
\Text(221,15)[]{$\gamma$}
\Photon(230,15)(275,15){3}{7}
\ArrowLine(310,50)(275,85)
\ArrowLine(275,15)(310,50)
\Text(375,50)[]{$h/H/A$}
\ArrowLine(275,85)(350,85)
\Text(360,85)[]{$\tau^-$}
\Text(280,30)[]{$\tau^-$}
\ArrowLine(350,15)(275,15)
\Text(360,15)[]{$\tau^+$}
\Text(280,70)[]{$\tau^+$}
\DashLine(310,49.96)(350,49.96){5}
\DashLine(310,50.04)(350,50.04){5}
\end{picture}
\end{center}
\caption{\it The process of $\tau\tau$ fusion  to Higgs bosons in 
 $\gamma\gamma$ collisions.} 
\label{fig:process}
\end{figure}

\vskip 3mm
{\bf 2.)} The formation of
the light and heavy $\Phi = h/H/A$ Higgs bosons   
in $\tau\tau$ fusion at 
a photon collider proceeds as shown in  Fig.\ref{fig:process}.
For the large-$\tan\beta$ case studied here, all
the Higgs bosons $\Phi$ decay almost exclusively [80 to 90\%]
to a pair of $b$ quarks. Therefore the final state consists of a pair of
$\tau$'s and a pair of resonant $b$ quark jets.

In the equivalent-particle approximation the process can be decomposed into
two consecutive steps: photon splittings to tau pairs, $\gamma \to \tau^+
\tau^-$, followed by the fusion process of two (almost on-shell) taus to the
Higgs bosons, $\tau^+ \tau^- \to \Phi$.
The cross section is given by the convolution of the fusion cross
section with the $\tau\tau$ luminosity in the colliding photon beams:
\begin{eqnarray}
   \sigma[\gamma \gamma \to \tau^+\tau^- \Phi ] 
        =\frac{\pi m_\tau^2}{2v^2 s} \, g^2_{\Phi\tau\tau} \times  
       2 \int_{\tau}^1 \frac{dz}{z}\,
          D^\tau_\gamma(z) D^\tau_\gamma(\tau / z )\,   
\label{gg2higgs}
\end{eqnarray}
$v$ is the Higgs vacuum expectation value, $v\simeq 246$~GeV; $\sqrt{s} =
E_{\gamma\gamma}$ is the c.m.~energy of the photons, and $\tau= M_\Phi^2/s$. 
The couplings 
$g_{\Phi\tau\tau}$ are normalized to the Standard Model Higgs coupling 
to a tau pair, $m_\tau/v$. For large $\tan\beta$, the couplings are given by
\begin{eqnarray}
g_{\Phi\tau\tau}& =& \tan\beta 
{\rm ~~~~~~~~~~ for ~} \Phi=A   \nonumber\\
g_{\Phi\tau\tau}&\simeq& \tan\beta 
{\rm ~~~~~~~~~~ for ~} \Phi=h,H
\end{eqnarray}
if the pseudoscalar mass parameter $M_A$ is sufficiently light in the case of
$h$, and sufficiently heavy in the case of $H$, {\it cf.}\ Ref.~\cite{R2} for
details. From the 
$\gamma \to \tau$ splitting function \cite{R7} $D_\gamma^\tau (z)$
the $\tau\tau$ luminosity 
\begin{eqnarray}
  F_{LL}(\tau) =    
     \left(\frac{\alpha}{2\pi}\right)^2 \,
 \log^2 \frac{M^2_\Phi}{m^2_\tau} \, \times \, 
                  [ 2 (1+2\tau)^2 \log\tau^{-1} -4(1-\tau)(1+3\tau)]
  \label{LL}
\end{eqnarray}
in the second part of Eq.(1) can easily be derived.
 
A rough estimate, based on the equivalent-particle approximation 
introduced in Eq.(1), 
shows the size
of the fusion cross section to be $\sim 8$ fb 
for the $\gamma\gamma$ c.m. energy
$E_{\gamma \gamma} = 600$ GeV 
and the Higgs parameters $M_{H/A} = 400$ GeV and $\tan\beta = 30$. For
an integrated luminosity of 200 fb$^{-1}$, about 3,000 events 
can be expected in
both $H$ and $A$ decay channels.  As a result, a statistical error of order
1\% can be predicted that compares favorably well with other methods 
\cite{R5,R6}. On the other hand, the light Higgs boson $h$ and, 
for moderate mass values, the heavy Higgs 
bosons $H,A$ can also be produced at lower energies, 
{\it e.g.}~$E_{\gamma\gamma}=400$~GeV.

In the same way the size of the cross section for the main background channel 
can be estimated: $\tau^+ \tau^-$ annihilation into a pair of $b$-quarks,
via $s$-channel $\gamma/Z$ exchanges.
As this mechanism is of higher order in the electroweak interactions, it is 
naturally small and strongly suppressed away from the $Z$ resonance region.
[The reverse process, annihilation of $b$'s to $\tau$'s, is very small
due to the fractional $b$ electric charge.]
A second background channel is associated with diffractive $\gamma \gamma \to
(\tau^+ \tau^-) (b\bar{b})$ events, the pairs scattering off each other by
Rutherford photon exchange. This diffractive background can be suppressed 
strongly by choosing proper cuts: the paired fermions 
in diffractive events travel 
preferentially parallel to the $\gamma$ axes and they carry small invariant 
mass, a topology quite different from the signal events. 
\begin{figure}[ht!]
\begin{center}
\epsfig{figure=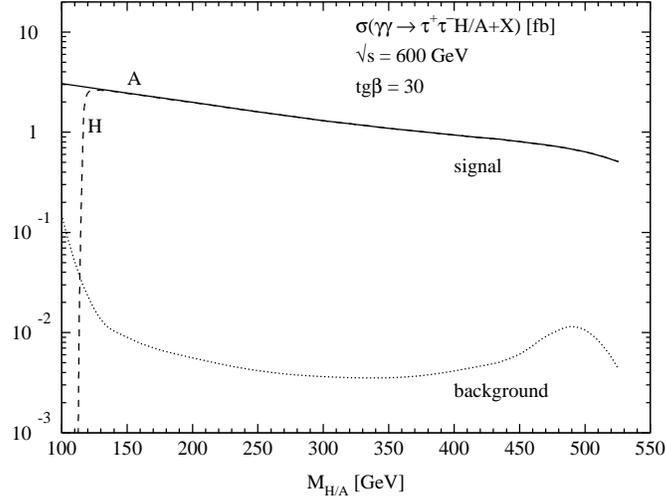,height=7.1cm}\\
\epsfig{figure=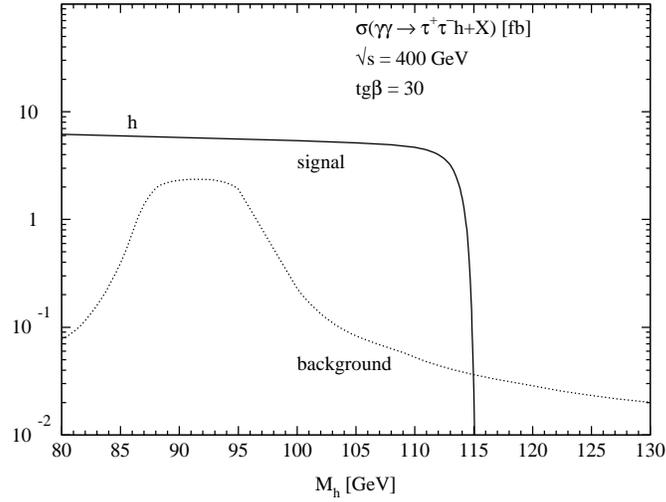,height=7.1cm}
\end{center}
\caption{\it The cross sections for the production of the $H/A$ (top) 
  and $h$ (bottom) Higgs bosons in the $\tau\tau$ fusion process 
  at a $\gamma\gamma$ collider for
  $\tan\beta=30$.  Also shown is the background cross section for experimental
  cuts as specified in the text.
  $\sqrt{s}$ denotes the $\gamma\gamma$ collider c.m.~energy, corresponding to
  approximately 80\% of the $e^\pm e^-$ linear collider energy.}
\label{fig:xsec}
\end{figure}
%

\vskip 3mm
{\bf 3.)} For an $e^\pm e^-$ collider c.m. energy of 800 GeV, 
the maximum of the $\gamma\gamma$ energy spectrum 
is close to 600 GeV. Adopting the detailed TESLA parameters, 
an integrated $\gamma\gamma$ 
luminosity of about 200 fb$^{-1}$ {\it per annum} can
be expected in the margin 20\% below the maximum $e^+e^-$ 
energy \cite{R10}. Similarly,
about 100 fb$^{-1}$ may be accumulated for a
$\gamma\gamma$ energy of 400 GeV at a 500 GeV $e^\pm e^-$ collider. 

In the numerical analysis the full set of diagrams for the signal processes
$\gamma \gamma \to \tau^+ \tau^- + \Phi [\to b\bar{b}]$ and all diagrams 
for the background processes $\gamma \gamma \to \tau^+ \tau^- b\bar{b}$, 
generated by means of {\tt CompHEP} \cite{R12}, are taken into account.
This set includes for the signal in particular the bremsstrahlung of the
Higgs bosons off the external $\tau$ legs.
  
The exact cross
sections for the signals of $H$ and $A$ Higgs-boson production
in the $\tau\tau$ fusion process, together with all the background 
processes, are presented in the top panel of Fig.~\ref{fig:xsec}.
The cuts on the final states have been chosen such that the
diffractive $\gamma$-exchange mechanism is suppressed sufficiently well: 
the invariant $b\bar{b}$ mass has been constrained to the
bracket $ \Delta=0.05 M_\Phi$, the taus are assumed visible and traveling in 
opposite directions to the beam axis with tau energies in excess of 5 GeV, 
and polar angles beyond 130 mrad to account for the shielding.
From the complementary bottom panel of Figure \ref{fig:xsec} it is clear that
$\tau\tau$ fusion to the light Higgs boson $h$ can also be exploited to measure
$\tan\beta$ for large values if the pseudoscalar mass is moderately small.\\

\begin{table}[ht!]
\begin{center}
\begin{tabular}{|c||c|cc||c|cccc|}
\hline\rule{0cm}{4mm}
 &\multicolumn{3}{|c||}{$E_{\gamma\gamma}=400$ GeV} 
 &\multicolumn{5}{|c|}{$E_{\gamma\gamma}=600$ GeV}\\
\hline
\rule{0cm}{5mm}$M_{\mathrm{Higgs}}$ & $A\oplus h$ &
\multicolumn{2}{|c||}{$A\oplus H$}& 
$A\oplus h$ &\multicolumn{4}{|c|}{$A\oplus H$}\\ 
 \phantom{i}[GeV]  
&  100  &  200  &  300  & 100   &  200  &  300  &  400  &  500  \\
\hline
\rule{0cm}{5mm}$\tan\beta$ & \scriptsize{I}   & \scriptsize{II}   & 
              \scriptsize{III} &
              \scriptsize{IV}  & \scriptsize{V}    & \scriptsize{VI}  &
              \scriptsize{VII} & \scriptsize{VIII}     \\
\hline\hline
\rule{0cm}{5mm}10  & 8.4\% & 10.7\% & 13.9\% & 8.0\% & 9.0\% & 
11.2\% & 13.2\% & 16.5\% \\
30  & 2.6\% & 3.5\% & 4.6\% & 2.4\% & 3.0\% & 3.7\% & 4.4\% & 5.3\% \\
50  & 1.5\% & 2.1\% & 2.7\% & 1.5\% & 1.8\% & 2.2\% & 2.6\% & 3.2\% \\
\hline
\end{tabular}
\caption{\it Relative errors $\Delta\tan\beta/\tan\beta$ on $\tan\beta$ 
in measurements for $\tan\beta=$ 10, 30 and 50, based on: 
combined $A\oplus h$ [I,IV] and $A\oplus H$ [II,III,V--VIII] production, 
assuming [$E_{\gamma\gamma}= 400$~GeV, ${\cal L}= 100$~fb$^{-1}$]
and  [$E_{\gamma\gamma}= 600$~GeV, 
${\cal L}= 200$~fb$^{-1}$]. Cuts and efficiencies are applied on the 
final--state $\tau$'s and $b$ jets as specified in the text.}
\label{tab:stat-error}
\end{center} 
\end{table}

The statistical accuracy with which large $\tan\beta$ values can be measured in
$\tau\tau$ fusion to Higgs bosons can be estimated from the predicted cross
sections and the assumed integrated luminosities. Efficiencies for $bb$ 
tagging, $\epsilon_{bb}$, and $\tau\tau$ tagging, $\epsilon_{\tau\tau}$, 
reduce the accuracy. For $\epsilon_{bb}\sim 0.7$ and 
$\epsilon_{\tau\tau}\sim 0.5$, for example \cite{R8}, the errors grow by a 
factor $1/\sqrt{ \epsilon_{bb}\epsilon_{\tau\tau}}\sim 1.7$. The expected 
errors for $h/H/A$ production are exemplified for three 
$\tan\beta$ values, $\tan\beta=10$, 30 and 50, in Table~\ref{tab:stat-error}. 
The integrated luminosities are chosen to be 200 fb$^{-1}$ for the high 
energy option and 100 fb$^{-1}$ for the low energy option \cite{R10}.
For $h$ production, the mass parameters are set to $M_A\sim 100$ GeV and
$M_h=100$ GeV; for the production of the heavy pseudoscalar $A$ the mass is
varied between 100 and 500 GeV. Results for scalar $H$ production are 
identical to pseudoscalar $A$ in the mass range above 120 GeV. 
The two channels $h$ and 
$A$, and $H$ and $A$ are combined in the overlapping mass ranges in which
the respective two states cannot be discriminated. In 
Table~\ref{tab:stat-error} we have presented the relative errors 
$\Delta\tan\beta/\tan\beta$. Since in the region of interest 
the $\tau\tau$ fusion cross sections are proportional to $\tan^2\beta$ and 
the background is small, the absolute errors $\Delta\tan\beta$ are nearly 
independent of $\tan\beta$, varying between 
\begin{eqnarray}
\Delta\tan\beta \simeq 0.9 {\;\;\;\rm and \;\;\;} 1.3          \label{deltat}
\end{eqnarray}
for Higgs mass values away from the kinematical limits.

It should be noted that away from the kinematical limits, the Higgs fusion
cross sections vary little with the $\gamma\gamma$ energy 
since the suppression of the parton subprocess with rising energy
is almost compensated by
the luminosity function. As a result, the smearing of the
$\gamma\gamma$ energy has a mild effect on the analysis presented here. 
Moreover, since the $\gamma\gamma$ machine-luminosity rises 
with the collider energy, the errors on $\tan\beta$ decrease
correspondingly.

\vskip 3mm
{\bf 4.)} $\tau\tau$ fusion to the heavy Higgs
bosons $H/A$ of the MSSM at a photon collider is a promising channel for
measuring the Higgs mixing parameter $\tan\beta$ at large values. Complemented
by $\tau\tau$ fusion to the light Higgs boson $h$ for moderately small values 
of the pseudoscalar Higgs boson mass $M_A$, the MSSM parameter range can nicely
be covered in all scenarios. This analysis compares favorably well with the
corresponding $b$-quark fusion process at the LHC \cite{R6}.  Moreover, the 
method can be applied readily for a large range of Higgs mass values and thus 
is competitive with complementary methods in the $e^+e^-$ mode of a linear 
collider \cite{R5}.

\vskip 3mm
{\bf References}


\begin{thebibliography}{99}
\bibitem{R2} 
M.~Spira and P.~M.~Zerwas,
Lectures at the  Schladming Winter School 1997 
[arXiv:hep-ph/9803257];
E.~Boos, A.~Djouadi, M.~M\"uhlleitner and A.~Vologdin,
\PRD \rm {\bf 66} (2002) 055004
[arXiv:hep-ph/0205160].

\bibitem{2a}
S.Y.~Choi, J.~Kalinowski, J.S.~Lee, M.M.~M\"uhlleitner, M.~Spira and P.M.~Zerwas,
arXiv:hep-ph/0404119 [to be published in \PLBB].

\bibitem{2b} M.~M.~M\"uhlleitner, M.~Kr\"amer, M.~Spira and P.~M.~Zerwas,
\PLB \rm {\bf 508} (2001) 311
[arXiv:hep-ph/0101083];
M.~M.~Velasco {\it et al.},
in {\it Proc. of the APS/DPF/DPB Summer Study on the Future of Particle 
Physics (Snowmass 2001)},
[arXiv:hep-ex/0111055];
D.~M.~Asner, J.~B.~Gronberg and J.~F.~Gunion, \PRD \rm {\bf 67} (2003)
035009.

\bibitem{2c} P.~Nie\.zurawski, A.~F.~\.Zarnecki and M.~Krawczyk,
arXiv:hep-ph/0307180, hep-ph/0307183, and hep-ph/0403138.

\bibitem{R7} M.~S.~Chen and P.~M.~Zerwas,
\PRD \rm {\bf 12} (1975) 187.

\bibitem{R5} 
 J.~F.~Gunion, {\it et al.}, in ``LHC/LC  Physics Document'', 2004.

\bibitem{R6} R.~Kinnunen, S.~Lehti, F.~Moortgat, S.~Nikitenko and M.~Spira,
 CMS--Note CMS AN 2003/014.

\bibitem{R10} 
B.~Badelek {\it et al.}  [ECFA/DESY Photon Collider Group], TESLA-TDR, 
Part VI, DESY 02-011
[arXiv:hep-ex/0108012];
E.~Boos {\it et al.},
\NIMA \rm {\bf 472} (2001) 100
[arXiv:hep-ph/0103090].

\bibitem{R12} A.~Pukhov {\it et al.}, CompHEP Collaboration, 
arXiv:hep-ph/9908288.

\bibitem{R8} K.~Desch, {\it private communication}.
\end{thebibliography}
\end{document}